\documentclass[twocolumn,showpacs,preprintnumbers,amsmath,amssymb]{revtex4}
\usepackage{dcolumn}
\usepackage{bm}
\usepackage{amssymb,amsmath, amsthm,epsfig}
\newcommand{\ra}{\rangle}
\newcommand{\oz}{\omega_z}
\newcommand{\la}{\langle}
\newcommand{\p}{\partial}
\newcommand{\kk}{{\bf k}}
\newcommand{\bk}{{\bf k}}

\begin{document}
\title{Resonance enhanced turbulent transport}
\author{Andrew PL Newton and Eun-jin Kim}
\affiliation{Department of Applied Mathematics, University of Sheffield, Sheffield, S3 7RH,U.K.}

\begin{abstract}
The effect of oscillatory shear flows on turbulent transport of passive scalar fields is studied by numerical computations based on the results provided by E. Kim [\emph{Physics of Plasmas}, {\bf 13}, 022308, 2006]. Turbulent diffusion is found to depend crucially on the competition between suppression due to shearing and enhancement due to resonances, depending on the characteristic time and length scales of shear flow and turbulence. Enhancements in transport occur for turbulence with finite memory time either due to Doppler and parametric resonances. Scalings of turbulence amplitude and transport are provided in different parameter spaces. The results suggest that oscillatory shear flows are not only less efficient in regulating turbulence, but also can enhance the value of turbulent diffusion, accelerating turbulent transport.
\end{abstract}

\maketitle

\vspace{0.1cm}

\section{Introduction}

Turbulent transport is an important process, by which various physical quantities
are rapidly mixed by the advection of turbulent fluid. Examples include the mixing
of chemical species, dissipation of magnetic fields, transport of angular momentum
in astrophysical/laboratory plasmas, the mixing of pollution in the terrestrial
atmosphere or even the transport of warm water in oceans. In fact, the effect of
turbulent transport in our lives is fundamental. Turbulent transport can however
have an unwelcoming consequence, such as anomalous heat/energy loss in laboratory
plasmas, leading to the degradation of plasma confinement. It is therefore crucial
to understand the physics of turbulent transport, especially some means of controlling it.

One of the promising mechanisms for quenching turbulence mixing is
flow shear suppression \cite{BURRELL,HAHM02,kimb}. The basic idea is
that shearing of turbulent eddies by flow shear leads to a cascade
of eddies into smaller and smaller scales down to the dissipative
scale where the molecular dissipation efficiently removes the
energy. That is, the shear flow rapidly generates small scales,
thereby enhancing the overall dissipation of turbulent eddies
which are responsible for turbulent transport. As a result, both
turbulence level and transport are reduced
\cite{kimb,KIM03,KIM04,kim05,KL}. In fact, there has been
accumulating experimental evidences for turbulence regulation by
flow shear in laboratory plasmas, which is now thought to be
indispensable for the formation of transport barriers \cite{BURRELL,HAHM02}. A crucial question in understanding the
formation of transport barrier in a variety of systems is thus, how
much of turbulent transport is reduced by a given flow shear
compared to the case in the absence of the flow shear. Quantitative
theoretical predictions for turbulent transport were provided in different types of turbulence models for steady \cite{KIM03,KL} or
random shear flows  \cite{KIM04,loconte}.

In comparison with a steady shear flow, turbulence regulation by
time-varying shear flow is more complex, with its efficiency
depending on characteristic time scales of the shear flow and
turbulence such as their frequencies, decorrelation times, shearing
rate,  etc.  \cite{KIM04,kim,HAHM}. For instance, if a shear flow
oscillates rapidly  with its frequency much larger than the
turbulent decorrelation rate, its shearing is too incoherent to have
any effect on turbulent transport. In contrast, if the flow shear is
sufficiently strong such that its shearing rate involves the fastest
time scale, it  can be considered to be steady and thus has a
similar effect on turbulent transport as a steady shear flow. For
instance, in this limit, turbulent mixing of passive scalar fields
is reduced inversely proportional to the rms shearing rate
$\Omega_m$ as $\propto 1/\Omega_m$ \cite{kim}. What happens in
between these two extreme limits is far less obvious. While
theoretical predictions in a scalar field model are provided by Kim \cite{kim}, they are given in terms of multiple integrals which
could be reduced to simple analytical forms only in the extreme
limits, yielding transparent scalings with $\Omega_m$ only in these
limits. Further quantitative study is thus necessary in order to
understand how efficiently an oscillatory shear flow controls
turbulent mixing in general.

It is important to emphasize that the aforementioned reduction in
turbulent transport by flow shear results from the modification of
the turbulence properties (e.g. enhanced dissipation of
turbulence/fluctuations) when the flow shear can have no influence on the 
mean field profile, for instance, when a mean field, being uniform
along the shear flow, varies only transverse to the shear flow. This
is relevant for the study of turbulent transport where main the interest
lies in the transport mediated by turbulence. In the case when a
mean field varies along the shear flow as well as transverse to it,
the shear flow has a direct influence on mean field since it can
distort the distribution of the mean field, thereby rapidly
generating small scales which then get efficiently damped by the
molecular diffusion [for example, see Refs. \cite{MOFFATT,CHIPPY,CHIPPY1}].
That is, the same shear distortion that enhances the dissipation of
turbulent eddies, leading to the reduction in turbulent transport,
can speed up the diffusion of mean field if it directly operates on the mean field. 
Therefore, in general, when shear flow can modify the
mean field as well as turbulence, the overall effect of the shear
flow on turbulent transport will be determined by the competition
between
 these two conflicting effects.

It is also important to note that
 propagating oscillatory shear flows (waves)
have often been invoked as a mechanism
for transport  \cite{knobloch}. Examples include internal
gravity waves which can transport momentum, mass, heat, etc in
astrophysical and geophysical systems [e.g. see, Ref \cite{Press2}]. The
transport by these waves is however a slow process compared to
turbulent transport as the former requires non-ideal effect such as
molecular dissipation. On the other hand, oscillatory shear flows
can potentially mediate fast transport by destabilizing the
equilibrium via parametric resonance \cite{nres,poul}. In the
presence of turbulence, the interaction between oscillatory shear
flows and turbulent flows could lead to similar parametric
resonance, as indicated by Kim \cite{kim}, significantly
contributing to the transport.
Without this resonance, the effect of oscillatory shear flows
on laminar diffusion can be opposite since the shear
enhanced mixing becomes negligible
as the oscillation frequency of the shear flow increases
above the rms shearing rate \cite{PARKER}.

The purpose of this paper is to perform a detailed quantitative study
of the effect of oscillatory shear flow on turbulent transport of
passive scalar fields. In particular, we derive the asymptotic
scalings of turbulence amplitude and transport with the rms shear
strength via numerical integrations of the theoretical results derived
by Kim \cite{kim}. We distinguish three different scaling regimes
depending on the characteristic time scales of turbulence and oscillatory
shear flow, identify the two types of resonances (Doppler and parametric
resonances), and then obtain asymptotic scalings with rms shearing rate
valid in each regime by numerical computations. We show that turbulent
diffusion depends crucially on the competition between suppression due
to shearing and enhancement due to the resonances, thereby suggesting
that oscillatory shear flows can enhance the value of turbulent diffusion,
accelerating turbulent mixing. We note that our numerical integration has
a great advantage of permitting a thorough parameter scan in parameter
space, which is not easily accessible by direct numerical computations
(e.g. in the limit of small dissipation). The remainder of the paper
is organized as follows. In \S II, we present the time-averaged turbulence
amplitude and transport in dimensionless form. In \S III, we identify the
three scaling regimes depending on characteristic time scales by simple
analytical examination. \S IV contains the results of numerical computations
and scalings of turbulence amplitude and transport with rms shearing rate.
Discussions and conclusion are provided in \S V.

\section{Governing equations}

We consider a passive scalar field model where a passive scalar $n$ is advected
by a given turbulent flow ${\bf u}$ and shear flow ${\bf U }_0$ while being
diffused by molecular dissipation $D$. By quasi-linear analysis, the fluctuating
scalar field $n'$ evolves according to the following equation:

 \begin{equation}
\{\partial_t+{\bf U }_0\cdot\nabla \}n'=- v_x\partial_x N_0+D\nabla^2 n'.
\label{eq1}
\end{equation}
Here, $N_0 = \langle n \rangle=N_0(x)$ is the large-scale component.
We assume that the shear flow is in the $y$ direction, varying
linearly in $x$, and is oscillatory in time of the form ${\bf U
}_0=-x\Omega_m\sin\left(\omega_z t \right){\bf y }$; $\omega_z$ and
$\Omega_m$ are the frequency and rms shearing rate of the
oscillating shear flow. Note that since both $N_0$ and ${\bf U}_0$
depend only on $x$ while ${\bf U}_0$ is in the $y$ direction, there is
no direct effect of the shear flow on the mean field $N_0$ (i.e.
${\bf U}_0 \cdot \nabla N_0 =0$). That is,
the shear flow influences the mean field only indirectly
through its effect on turbulence.

The advection by a linear shear flow $U(x,t)= -x \Omega(t)$ results
in the distortion of an eddy (i.e., wind-up), and its effect can be
non-perturbatively captured by employing a time-dependent wavenumber
$k_x(t)$ with the following transformation for $n'$:
\begin{equation}
n'({\bf x},t) = {\tilde n}({\bf k},t) \exp{\{i(k_x(t) x + k_y y)\}}\,,
\label{eq2}
\end{equation}
and similarly for $v_x$, with $k_x$ satisfying an eikonal equation
\begin{equation}
\partial_t k_x(t) = k_y \Omega(t)\,. \label{eq3}
\end{equation}
It is worth noting that Eq.\ (\ref{eq3}) clearly shows that the
shear flow has no influence on the mean field ($k_y=0$ mode) of our
interest. The solution to Eq.\ (\ref{eq1}) can then be expressed as:
\begin{eqnarray}
{\tilde n}({\kk},t) &=-\p_x n_0 & \int^t_{-\infty} dt_1 d^2k_1 {\hat
g}(\kk,t;\kk_1,t_1) e^{-DQ(t,t_1)} {\tilde v}_x(\kk_1, t_1)\,.
\label{eq6}
\end{eqnarray}
Here, $Q(t,t_1) =  \int^t_{t_1} dt' [k_x^2(t') + k_y^2]$, and ${\hat
g}$ is the Green's function given by
\begin{equation}
{\hat g}({\kk},t;k_1,t_1) = \delta(k_y-k_{1y}) \delta \left[k_{x}
-k_{1x} - k_{1y} \int^t_{t_1} dt'\Omega(t')\right]\,. \label{eq7}
\end{equation}
The overall effect of enhanced dissipation due to shearing is
embedded in the time integral of $k^2$ in $DQ$ in Eq.\ (\ref{eq6}).
The detailed form of $Q$ for $\Omega(t) = -\Omega_m \sin{\omega_z t}$ was
provided by Kim \cite{kim}, to which the readers are referred.
We here emphasize that  $Q$ grows at most linearly in
time for the oscillatory shear flow as $k_x$ 
oscillates in time (see Eq.\ (\ref{eq3})),
with its shearing becoming effective 
only when it operates coherently before the oscillatory
zonal flows change
shearing direction, i.e. when $\Omega_m/\oz>1$ \cite{kim}. 
This is  distinctively  different
from the behaviour of $Q$ in the case of mean shear flows or random
zonal flows: (i) in
the case of mean shear flows \cite{KIM03}, $k_x$ grows linearly in time with $Q
\propto t^3$; (ii) in the case of random zonal flows \cite{KIM04} $k_x$ can be considered as a
random process with the time average of $k_x^2 \propto t$ on a long,
diffusive time scale, and thus $Q \propto t^2$. 

The flux and amplitude are obtained by assuming
that the statistics of the turbulent flow $v_x$
are spatially homogeneous and temporally stationary with the
following correlation function:
\begin{equation}
\langle {\tilde v}_x(\bk_1,t_1) {\tilde v}_x(\bk_2,t_2) \rangle = (2
\pi)^2 \delta(\bk_1+\bk_2) {\phi} (\bk_2, t_2-t_1)\,, \label{eq44}
\end{equation}
where $\phi$ is the correlation function of $v_x$ in Fourier space.
Further, the random turbulent flow is taken to have 
characteristic frequency $\omega$ and correlation time
$\tau_c=1/\gamma$. Specifically, $\psi$ is taken to have Lorentzian
frequency spectrum centered around $\omega$ with width $\gamma$ as
$\phi({\bf k}_2,t_2-t_1) = \psi({\bf k}_2) \int (d\omega'/\pi)
 e^{-i\omega'(t_2-t_1)}
\gamma/[(\omega'-\omega)^2+\gamma^2]$ with $\omega>\gamma$, where
c.c. denotes complex conjugate. It can be readily shown that the
velocity amplitude is related to the power spectrum $\psi$ as  $\la
v_x^2 \ra = \int d^2 k \psi(\kk)/(2\pi)^2$. Since $k_x \ne 0$ modes
are
 generated by the shear [see, Eq.\ (\ref{eq3})], $\psi$ is
assumed to be dominated by modes with $k_{x} \ll k_{y}$ for
simplicity. Then, 
the flux and amplitude of fluctuation of scalar
fields, averaged over one oscillation of the shear flow
($2\pi/\omega_z$) as well as over the statistics of the turbulence
follow from Eqs.\ (\ref{eq6}) and (\ref{eq44})
(see Eqs. (9)-(10) in  \cite{kim}) and can be expressed
in the following dimensionless form:

\begin{equation}\label{flux}
\begin{split}
\la n'v_x\ra_t = -\frac{\p_x n_0}{(2 \pi)^2} \int d^2 k
\psi(\kk) \int_0^{2\pi}\int_0^{\tau}d\tau d\tau_1
A(\beta,\overline{\omega},\tau,\tau_1)\\ \times
e^{-\overline{\gamma}(\tau-\tau_1)-\tau_{D}^{-1}B(\alpha,\tau,\tau_1)},
\end{split}
\end{equation}
and

\begin{equation}\label{amplitude}
\begin{split}
\la n'^2\ra_t= \frac{(\p_x n_0)^2} {(2 \pi)^2 } \int d^2 k
\psi(\kk) \int_0^{2\pi}\int_{0}^{\tau}\int_{0}^{\tau}d\tau
d\tau_2d\tau_1 A(\beta,\overline{\omega},\tau_1,\tau_2)\\ \times
e^{-\overline{\gamma}(\tau_1-\tau_2)-\tau_{D}^{-1}
\left(B(\alpha,\tau,\tau_1)+B(\alpha,\tau,\tau_2)\right)}.
\end{split}
\end{equation}
Here,

\begin{equation}\label{A}
A(\beta,\overline{\omega},\tau,\tau_1)=
\cos \{\overline{\beta}\alpha(\cos(\tau)-\cos(\tau_1))
-\overline{\omega}(\tau-\tau_1)\},
\nonumber
\end{equation}
and

\begin{equation}\label{B}
\begin{split}
B(\alpha,\tau,\tau_1)=(\tau-\tau_1)\{1+\alpha^2(1+\frac{1}{2}\cos{2\tau_1})\\
\mbox{        }
+\frac{\alpha^2}{4}\left(\sin{2\tau}+3\sin{2\tau_1}-8\cos{\tau_1}\sin{\tau}\right)\}.
\nonumber
\end{split}
\end{equation}
The dimensionless variables in Eqs. (\ref{flux}) and (\ref{amplitude})
are defined using $\omega_z^{-1}$ as a unit of time
(i.e., $\tau=\omega_z t$) as follows:

\begin{description}
\item [-]$\alpha = \frac{\Omega_m}{\omega_z} \propto$ rms shearing rate
\item [-]$\overline{\gamma} = \frac{\gamma}{\omega_z} =
\frac{1}{\tau_c \omega_z} \propto$ decorrelation rate of turbulence
\item [-]$\overline{\beta}= kx \propto$ scale separation between the mean and fluctuations
\item [-]$\overline\omega=\frac{\omega}{\omega_{z}} \propto$ frequency of turbulence
\item [-]$\tau_D = \frac{\omega_z}{D k^{2}} \propto $ molecular diffusion time scale
\end{description}
Note that $k=k_y$ in $\overline{\beta}$ is the typical wavenumber of
turbulence in the $y$ direction while $k_x(t) = k_x(t_1) + k_y
\int^t_{t_1} dt' \Omega_m \sin (\omega_z t')$ evolves in time due to
shearing [see Eq. (\ref{eq3})].

It is important to note that the flux can be expressed by using
turbulent diffusivity $D_T$ as $\la n'v_x\ra_t  = - D_T \partial_x
N_0$ in homogeneous and stationary turbulence. In the case of
inhomogeneous turbulence (e.g., due to the background density
stratification or gradient in turbulence intensity), non-diffusive
flux can appear in the flux [for example see Ref \cite{KICHATINOV}].
Further, $D_T$ is likely to be positive, especially for a
short-correlated turbulence or for weak turbulence quenched by
strong shear, where a quasi-linear analysis becomes exact. Thus, in
our case, the turbulent diffusivity increases the dissipation rate of
large-scale component $N_0$ from the molecular value $D$ to a larger
value $D + D_T$. In the absence of shear flow, or equivalently, in
the limit $\alpha \to 0$, turbulent transport due to background
turbulence is fast with the typical value of $D_T \sim u l$, where
$u$ and $l$ are the characteristic velocity and length scale of
turbulence. In the following sections, we will investigate how the
turbulent transport and turbulence amplitude are affected by an 
oscillatory shear flow via numerical integration of Eqs.
(\ref{flux}) and (\ref{amplitude}) by varying the values of
$\alpha$, $\overline{\gamma}$, $\overline{\beta}$, $\overline\omega$
and $\tau_D$. In all cases, the molecular dissipation time is
assumed to be large with $\tau_D \gg 1$.

\section{Regimes of different scalings}

In this section, we identify the three regimes of different scalings depending on the
order of the relevant time scales.

\subsection{Short correlation time/period}
In the limit of short correlation time $\tau_c$ of turbulence
($\overline\gamma\rightarrow \infty$), double integrals of
$e^{-\overline\gamma(\tau-\tau_1)}$ in Eqs. (1) and (2) indicate that
both the flux and the amplitude decay $\propto\overline\gamma^{-1}$.
Similarly, in the large frequency limit ($\overline{\omega}\rightarrow\infty$),
both quantities scale as  $\propto\overline{\omega}^{-2}$. In this regime,
there is no possibility of resonance between shear flow and turbulence,
with turbulence level and transport decreasing as either $\overline\gamma$
or $\overline{\omega}$ increases. This is because the time scales for
the shear flow ($\tau_z = \omega_z^{-1}$ and $\Omega_m^{-1}$) are far
too large in comparison with the small correlation time ($\tau_c = \gamma^{-1})$
or characteristic period ($\omega^{-1}$) of turbulence.
Such a turbulent fluid is not able to recognize the subtle changes induced
by the shear flow over its correlation time or period and is thus not
affected by them. For instance, $\overline{\gamma}\gg1$ implies that
$\gamma \gg \omega_z \iff \tau_c \ll \tau_z$, i.e., the turbulence decorrelates
too rapidly to be influenced by the oscillation of the shear flow.

\subsection{Medium correlation time}
When turbulence has a longer correlation time with a finite value
of $1< \overline{\gamma}\ll\infty$ in the limit of small dissipation
$\tau_D \gg 1$, the flux in Eq. (\ref{flux}) and amplitude in
Eq. (\ref{amplitude}) take their maximum values when
$A(\beta,\overline{\omega},\tau,\tau_1)=1$, i.e. when

\begin{equation}\label{nit}
\overline{\beta}\alpha
\left[\cos(\tau)-\cos(\tau_1)\right]-\overline{\omega}(\tau-\tau_1)=0.
\end{equation}
To obtain the condition for the occurrence of maxima, Eq. \ref{nit},
we note that for $\overline{\gamma}>1$, most of the contributions
to the flux and amplitude in Eqs. (\ref{flux}) and
(\ref{amplitude}) come from the time integral for $|\tau - \tau_1| < 1$.
We can thus express
$\cos$ in terms of $\sin$ in Eq. (\ref{nit}), expand $\sin$ as a
Taylor series, and then factorize the first two leading order terms in
the series for $|\tau-\tau_1|\ll 1$ to obtain the following condition:

\begin{equation}\label{max}
\overline{\omega}=\overline{\beta}\alpha c
\iff \omega/k - c\sqrt{\la  U_0 ^{2}\ra_t  }= 0.
\end{equation}
Here, $c=|f(\tau,\tau_1)|<1$ is a constant of less than unity, whose exact
value depends on other parameter values.
Since $\omega/k=\omega/k_y$ is proportional to the phase speed of turbulence,
it is apparent that resonance occurs when the phase speed is approximately
equal to the rms velocity of the shear flow. This is equivalent to the Doppler
shifted frequency $\omega_d$ with the rms velocity of the shear flow being zero:

\begin{eqnarray}\label{doppler}
 \omega_d
&\approx&\omega-c{k_y }\sqrt{\langle U_0(x,t)^2\rangle_t}
= \omega -c \overline{\beta} \Omega_m \approx 0.\nonumber\\
\end{eqnarray}
Note that a similar resonance condition in terms of rms velocity also
holds in the case of random shear flow \cite{KIM04} while it was overlooked
in the analysis of an oscillatory shear flow \cite{kim}. In the case of a
steady shear flow $U_0 {\hat y}$, \cite{KIM03}, the resonance condition (\ref{doppler}) becomes exact as $\omega_d = \omega - k_y U_0 = 0$.

\subsection{Large correlation time}
When the turbulence correlation time is even longer with the value
$\overline{\gamma} \ll 1$, the flux and amplitude in Eqs. (\ref{flux})
and (\ref{amplitude}) can have considerable contributions from the time
integrals for all values of $|\tau - \tau_1|$, leading to the possibility
of the parametric type of resonance when $\omega=n\omega_z$ for integer $n$.
As indicated by Kim \cite{kim}, the parametric resonance can be found by
expanding $e^{i\overline{\beta} \alpha \cos{\tau}}$ in terms of Bessel functions.
We confirm numerically that Eqs. (\ref{flux}) and (\ref{amplitude}) do
take maximum values when $\omega=n\omega$ for integer $n$ in this case.

\section{Numerical Results}
In this section, we provide the quantitative predictions for turbulence
level and turbulent transport in different regimes identified in \S III,
by  numerical computations. The numerical accuracy was checked by benchmarking the code against an exactly integrable function which replicates the behaviour of our actual flux and amplitude. The numerical error  was found to be within $0.0001$ and $0.001$ for the flux and amplitude respectively.

\subsection{Medium correlation time}
We consider the medium correlation time in the limit of small molecular
dissipation (large $\tau_D$). Specifically, we take the value
$\overline{\gamma}=5$ and $\tau_D=100$.  The scale separation is chosen to
be $\overline{\beta}=10$. By using these parameter values and by varying $\alpha$ and
$\overline{\omega}$ within the ranges of $0\le \alpha \le 10$ and
$0 \le{\overline{\omega}} \le 100$, we perform numerical integrations of
Eqs. (\ref{flux}) and (\ref{amplitude}) to obtain the scalings of turbulence
amplitude and transport with the shearing rate $\Omega_m$. These parameter
values remain fixed unless stated otherwise.
\begin{figure}
   \begin{center}
   \includegraphics[width=8.0cm]{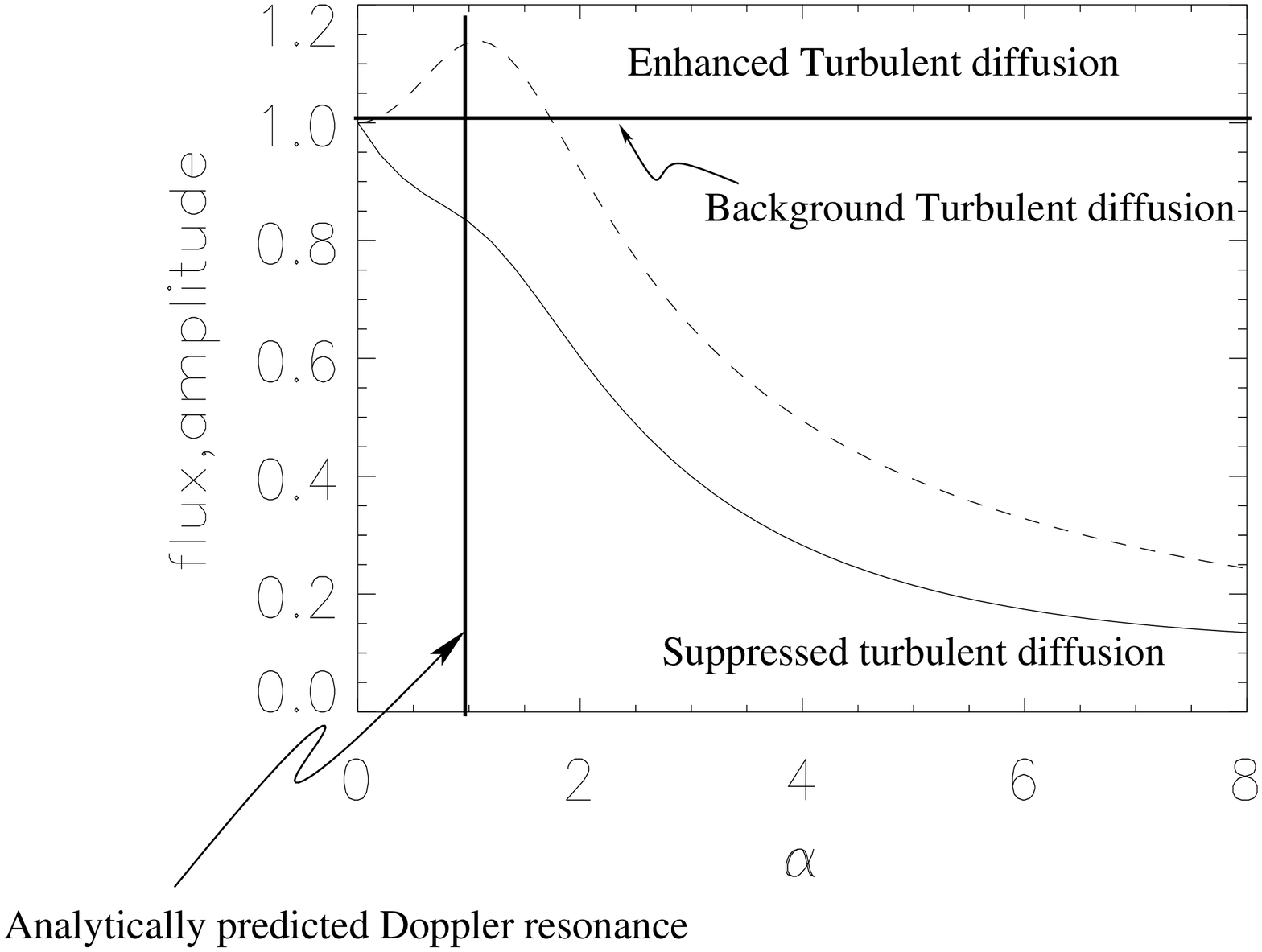}
   \caption{\label{fig1}Flux (dotted line) and amplitude (solid line) normalized to the $\alpha=0$ case}
Figure 1 caption: Flux (dotted line) and amplitude (solid line) normalized to the $\alpha=0$ case
   \end{center}
\end{figure}

\subsubsection{For a given $\overline{\omega}$}
For a given turbulence frequency $\overline{\omega}$, we normalize turbulence amplitude and transport by using values provided by background turbulence in the absence of shear flow (i.e. for $\alpha =0$). Specifically, we choose the frequency ratio $\overline{\omega}=10$ and plot in Fig. 1 the flux and amplitude normalized to the $\alpha=0$ case as a function of the rms shearing rate within the range of $\alpha\in[0,10]$. We note that as a consequence of our normalization, the value of unity for the flux and amplitude at $\alpha=0$ corresponds to that given by background turbulence in the absence of shear flow, with smaller or larger values describing the suppression or enhancement of turbulent diffusion, respectively. It is clear from Fig. 1 that as $\alpha$ increases from zero, the flux increases, indicating the enhancement relative to the case without a shear flow until it reaches its maximum, roughly at the Doppler resonance point, as noted in \S III B.  As the shearing rate increases further beyond this resonance point, the flux becomes quenched by strong shear. Similar behaviour, correlating well with the flux maximum, is also found in the amplitude although it is far less obvious.

\begin{figure}
   \centering
   \includegraphics[width=8.0cm]{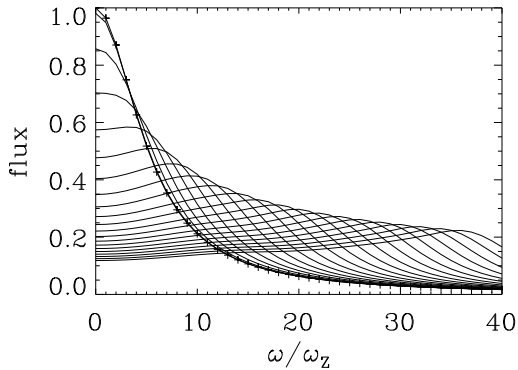}
   \caption{\label{fig2} Flux as a function of $\overline{\omega} = \omega/\omega_z$ with $\alpha$ increasing from top to bottom: top line no shear, bottom line strong shear}
\end{figure}
\subsubsection{Flux for different $\overline{\omega}$}
In the absence of shear flow ($\alpha=0$), the flux due to background turbulence depends on the  turbulence frequency. Thus, in order to examine the flux for different values of $\overline{\omega}$ whilst varying $\alpha$, we normalize the flux by its value in the case without a shear flow ($\alpha=0$) and for zero turbulence frequency $\overline\omega=0$. With this normalization, we plot the flux as a function of the frequency ratio $\overline{\omega}$ for different $\alpha\in\{0,0.2,...,4\}$ in Fig. 2. Fig. 3 shows the corresponding log plot. From Figs. 2 and 3, one can see the two regimes of clearly different scalings. First, in the limit of large turbulence frequency, the flux always eventually decays $\propto\overline{\omega}^{-1.96\pm0.06}$. This is consistent with the analytically predicted behavior $\propto\overline{\omega}^{-2}$ in \S III A. On the other hand, the maximum value of the flux at the Doppler resonance point decays $\propto\overline{\omega}^{-0.486\pm0.01}$, much slower with increasing $\overline{\omega}$. Fig. 4 shows similar behavior as rms shearing rate $\alpha$ increases. Specifically, in the strong shear limit, flux eventually decays proportionally to $\alpha^{-1.015\pm 0.01} \propto \Omega_m^{-1.015\pm 0.01}$, in agreement with Kim \cite{kim} while the resonant flux decays proportionally to $\alpha^{-0.534\pm0.01} \propto \Omega_m^{-0.534\pm0.01}$. Therefore, the scaling of the flux with $\Omega_m$ is weaker for maximum resonance flux with roughly $-1/2$ power-law dependence. Furthermore, the absolute value of the flux at the resonance is larger than that due to background turbulence, manifesting the resonance enhanced transport. In other words, oscillatory shear flow can enhance the overall transport above the value given by background turbulence due to resonance while it quenches transport for sufficiently strong rms shear. The resonance enhanced transport becomes more apparent if the flux is normalized to the case $\alpha=0$ case for all possible values of turbulence frequency $\omega$ in which case turbulent diffusion scales as $\propto \alpha^{1.420\pm0.1} \propto \Omega_m^{1.420\pm0.1}$. That is, turbulent diffusion increases because of the oscillatory shear flow!

\begin{figure}
   \centering
   \includegraphics[width=8.0cm]{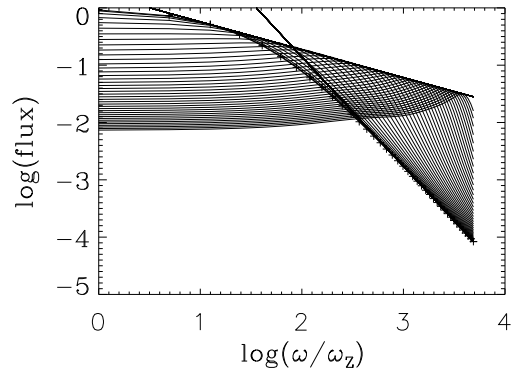}
   \caption{\label{fig3} Logarithmic plot of Fig \ref{fig2}}
\end{figure}
\begin{figure}
   \centering
   \includegraphics[width=8.0cm]{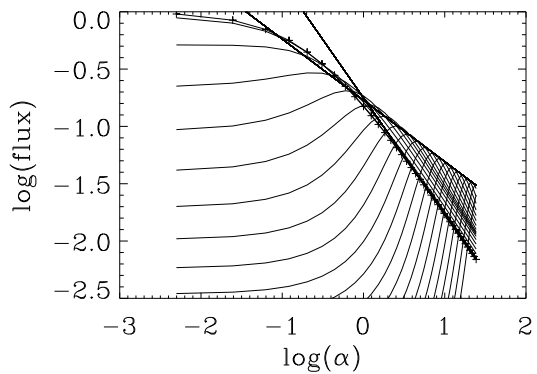}
   \caption{\label{fig4} Flux as a function of $\alpha$ for $\overline{\omega} = \omega/\omega_z$  with $\overline{\omega}$ increasing from top (random turbulence) to bottom (wave like turbulence)}
\end{figure}

\subsubsection{Amplitude for different $\overline{\omega}$}
Fig. 5 shows turbulence amplitude as a function of $\overline{\omega}\in [0,40]$ for different values of $\alpha\in\{0,0.2,...,4\}$, again normalized to the $\alpha=\omega=0$ case.
In the limit of large turbulence frequency, the amplitude decreases as $\propto\overline{\omega}^{-1.901\pm0.1}$ while the resonant amplitude decays less rapidly with $\overline{\omega}$ as  $\propto\overline{\omega}^{-0.476\pm0.05}$. On the other hand, Fig. 6 is a log plot of the amplitude as a function of $\alpha$, showing the two regimes with clear scalings with $\alpha$. In the strong shear limit,  amplitude $\propto\alpha^{-1.402\pm0.05}\propto\Omega^{-1.402\pm0.05}$ while the resonant amplitude $\propto\alpha^{-0.533\pm0.05}\Omega^{-0.533\pm0.05}$. Therefore, maximum amplitude has a slower decrease as shearing rate increases compared to asymptotic value of amplitude in the limit of strong shear.
\begin{figure}
   \centering
   \includegraphics[width=8.0cm]{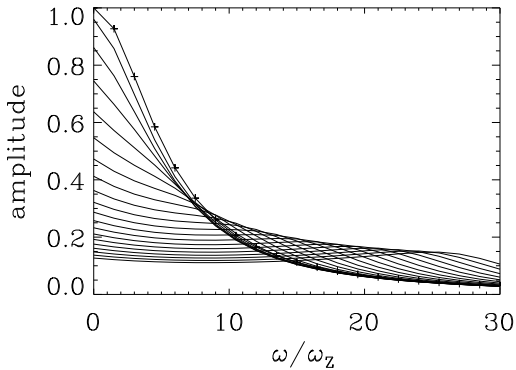}
   \caption{\label{fig5} Turbulence amplitude as a function of $\overline{\omega} = \omega/\omega_z$ with $\alpha$ increasing from top (no shear) to bottom (strong shear)}
\end{figure}
\begin{figure}
   \centering
   \includegraphics[width=8.0cm]{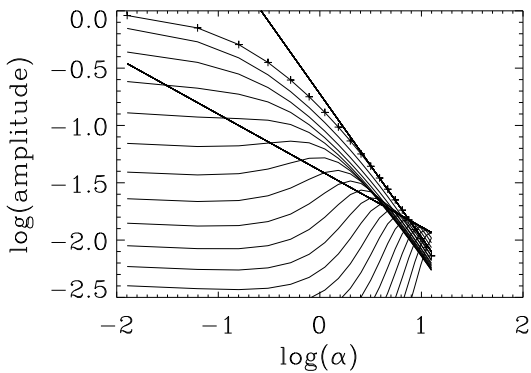}
   \caption{\label{fig6} Turbulence amplitude as a function of $\alpha$ for different values of $\overline{\omega}$ with $\overline{\omega}$ increasing from (random turbulence) to bottom (wave like turbulence)} 
\end{figure}
\subsubsection{Cross-phase results}
The flux $\langle n' v_x \rangle$ involves not only the phase relation between $n'$ and $v_x$ but also turbulence amplitude. We quantify the flux due to phase shift between $n'$ and $v_x$ only, independent of turbulence level, by defining the normalized flux (the so-called cross-phase) $\cos\theta$ as
\begin{equation}
\cos\theta=\frac{\la n' v_{x} \ra}{\sqrt{\la n'^2\ra \la v_x^2 \ra}}.\nonumber\\
\end{equation}
By using the results obtained in the previous subsections, we obtain the scaling of cross-phase with $\overline{\omega}$ in the large frequency $\overline{\omega}$ limit and at the Doppler resonance points as:

\begin{eqnarray}
\cos \theta &\propto&\overline{\omega}^{-1.009\pm0.035}, \nonumber\\
\cos \theta &\propto&\overline{\omega}^{-0.248\pm0.035}, \nonumber
\end{eqnarray}
respectively.
Similarly,  in the strong shear limit and at the resonance points, the cross-phases scale with $\alpha$ ($ \propto \Omega_m$) as

\begin{eqnarray}
\cos \theta &\propto&\Omega_m^{-0.315\pm0.035},\label{cross}\\
\cos \theta &\propto&\Omega_m^{-0.267\pm0.11},\nonumber
\end{eqnarray}
respectively. The integration accuracy is always less than  $ 0.0001$ and $ 0.001$ for the flux and amplitude accordingly, consequently errors in the scalings were found and related to the cross-phase.

In comparison with a steady shear flow case where $\cos \theta \propto \Omega^{-1/6}$ \cite{KIM03,kimb}, $\cos{\theta}$ in Eq. (\ref{cross}) have stronger dependence on $\Omega_m$, suggesting that oscillatory shear flow is more efficient than a steady shear flow at reducing normalized transport by affecting the phase relation between $n'$ and $v_x$. In particular, $\cos\theta\propto\Omega_m^{-0.315\pm0.035}$ in the strong shear limit, with its value decreasing rapidly compared to the steady shear flow case \cite{KIM03}. Therefore, oscillatory shear flows are more efficient in reducing normalized flux by quenching flux more than amplitude.

\begin{figure}
   \centering
   \includegraphics[width=8.0cm]{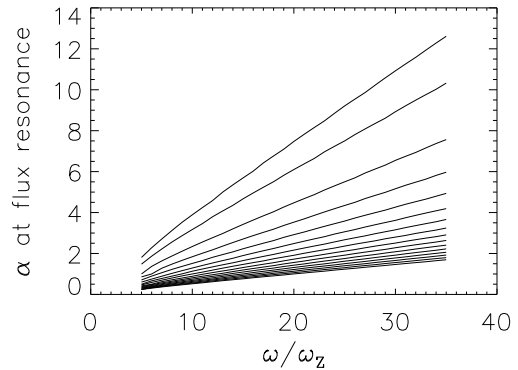}
   \caption{\label{fig7} Relationship between $[\alpha,\overline{\omega}]$ for resonant points with $\overline{\beta}$ increasing from top to bottom}
\end{figure}
\begin{figure}
   \centering
   \includegraphics[width=8.0cm]{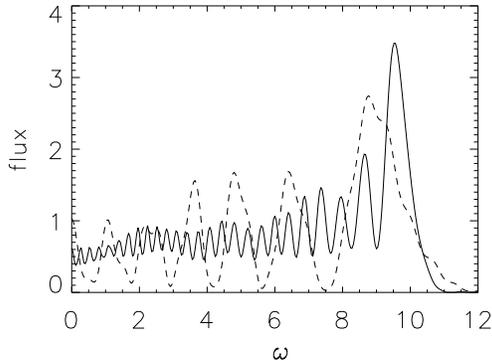}
   \caption{\label{fig8} Flux as a function of $\omega$ for $\omega_z=0.5$ (dotted line) and $\omega_z=0.125$ (solid line) when
$\overline{\gamma}=0.01$, showing multiple harmonic resonance in addition to Doppler resonance}
\end{figure}
\subsubsection{Behaviour of the Doppler resonance}
Figure 6 shows us the relation between the rms shearing rate of the oscillatory shear flow and turbulence frequency at the resonance for different values of the scale separation. In each case the relationship appears to be linear, with its proportionality decreasing as scale separation increases. Specifically, the relationship between these quantities at the peak resonance is found to be  $\overline{\omega}\propto\overline{\beta}^{-1.001\pm0.01}\alpha$ for the flux and $\overline{\omega}\propto\overline{\beta}^{-0.944\pm0.05}\alpha$ for the amplitude. This is consistent with analytical estimate in \S III within the numerical error caused by finite computing time.

\subsection{Long correlation time limit}
For sufficiently long correlation time of turbulence, parametric type resonance between oscillatory shear flow and turbulence can occur where $\omega  = n \omega_z$ for integer $n$, as discussed previously in \S III C (see also  \cite{kim}). In order to explore this resonance, we choose sufficiently small $\overline{\gamma}$ with a specific value of  $\overline{\gamma}=0.01$ whilst taking the dissipation time to be $\tau_D=10^3$. Furthermore, we change our non dimensional variables so that $\omega$ and $\omega_z$ are the only free parameters and alter the time average from being over $\left[0,2\pi\right]$ to $\left[0,2\pi/\omega_z\right]$.

Figure 8 is the plot of the flux as a function of $\omega$ for $\omega_z=0.5$ (dotted line) and $\omega_z=0.125$ (solid line) for the fixed values $\overline{\gamma}=0.01$ and $\tau_D=10^3$, and shows that the resonance peaks are equally spaced and increase in intensity until it reaches the transition phase into the Doppler resonance. Interestingly, the flux at Doppler resonance exceeds that due to parametric resonances. Beyond the Doppler resonance point, the flux decays rapidly to zero as shearing rate increases. Figure 9 shows how reducing the decorrelation rate (to $\overline{\gamma}= 10^{-4}$) and molecular dissipation (to $\tau_D=10^4$)  acts to amplify the height of these peaks as further harmonics appear. The height of resonance peaks would blow up in the limit of infinite memory time of turbulence (i.e. no stochasticity $\overline{\gamma} \to 0$) and no molecular dissipation (i.e. $\tau_D \to \infty$), leading to parametric instability  \cite{nres}. Parametric resonance can be physically understood since the equilibrium provided by a large-scale shear flow with frequency $\omega_z$ requires fluctuations to be invariant under the time translation by $1/\omega_z$, thereby supporting the excitation of modes of frequencies of $\omega= n \omega_z$. It is worth noting that a classical example of parametric resonance is a vertically oscillating pendulum \cite{Stoker}. The pendulum sways side to side and the frequency of these oscillations are integer multiples of the driving frequency, depending on the magnitude of the oscillations. For instance, the case when the driving frequency and the pendulum oscillation frequency match is simply explained by the driving frequency trying to force the pendulum into step.

\begin{figure}
   \centering
   \includegraphics[width=8.0cm]{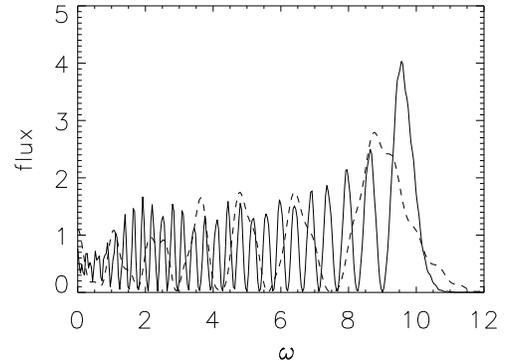}
   \caption{\label{fig9} Flux as a function of $\omega$  for $\omega_z=0.5$ (dotted line) and $\omega_z=0.125$ (solid line) when $\overline{\gamma}=10^{-4}$,
showing more pronounced resonant peaks than Fig. 8}
\end{figure}
\section{Conclusion}

We have performed a detailed study of the effect of oscillatory shear flow on turbulent transport and amplitude in passive scalar field model. Specifically, we have identified the three distinct scaling regimes with rms shearing rate, depending on the value of turbulence decorrelation time.
\begin{itemize}
\item Short correlation time limit ($\overline{\gamma}\rightarrow\infty$): there is no resonance due to too rapid change in turbulence characteristics, with the flux decaying as $\overline\gamma^{-1}$.
\item Medium correlation time ($1<\overline{\gamma}\ll\infty$):  turbulent transport is enhanced around the rms Doppler resonance where the wave phase speed matches the rms velocity of oscillatory shear. Increasing either the shear or the characteristic frequency from Doppler resonance point acts to quench turbulent transport. Our numerical results with $\overline\gamma=5$ show that if the flux is normalized to the case of $\alpha=0$ and $\overline{\omega}=0$, the flux, thereby turbulent diffusion, scales as, $\propto \Omega_m^{-0.5343}$ at resonance points. The dependence on $\Omega_m$ at resonance point is thus weaker than that in the strong shear limit ($\propto \Omega_m^{-1.015}$). Furthermore, the value of flux at resonance point is larger than that in the absence of the shear flow, manifesting the enhancement of turbulent transport due to the oscillating shear flow. If the flux is normalized to the case $\alpha = 0$, the flux scales as $\propto \Omega_m^{0.831}$ for all possible values of turbulence frequency $\omega$, highlighting the increase in turbulent diffusion due to oscillatory shear flow.
\item Long correlation time limit  ($0<\overline{\gamma} \ll 1$): there is parametric (harmonic) resonance of the form $\omega=n\omega_z\mbox{ where }n\in\mathbb N$ in addition to Doppler resonance. At these resonance points, the numerically computed flux and amplitude show significant enhancements, with their maximum values provided by Doppler resonance peaks.
\end{itemize}

These results are summarised in Table 1. Our results suggest that oscillatory shear flows can enhance the
value of turbulent diffusion, speeding up turbulent mixing,
depending on the characteristics of shear flow and turbulence.
Enhancement of turbulent mixing can then have either welcoming or
unwelcoming consequences in turbulent mixing. A typical example of
the latter can be found in laboratory plasmas where the confinement
is a critical issue. These results thus suggest that for the
understanding and predictive modelling of turbulent transport in
plasmas, it is necessary to determine frequency and power spectra of
shear flows and turbulence. Our results have implications for
turbulent mixing in many other fields such as geophysics,
oceanography, atmospheric physics, solar physics, and
magnetohydrodynamics where shear flows and turbulence are main
players in transport. In particular, similar results obtained in the
passive scalar fields model are expected to be valid for the
transport of magnetic fields in the 2D magnetohydrodynamic
turbulence as long as the backreaction of magnetic fields is
negligible (i.e. in the kinematic regime). An interesting question
is then what happens to the transport/diffusion of large-scale
magnetic fields when the backreaction is sufficiently strong to
modify the characteristics of turbulence. In particular, it would be
interesting to study whether oscillatory shear flows can weaken the
severe quenching in the amplification of magnetic fields (the
so-called dynamos) in 3D and their diffusion rate in 2D due to the
magnetic backreactions. Furthermore, the backreactions of turbulence
on shear flows will also have an importance consequence on the
evolution of shear flows, for instance, leading to complex temporal
and spatial dynamics \cite{CHIPPY1}, thereby dynamically determining
the frequency and power spectra of shear flows, which have been
assumed to be given in this paper and the previous works
 \cite{KIM03,KIM04,kim,KL}. These problems will be addressed in
future publications.


\vskip0.5cm
\noindent
{\bf Acknowledgements}\
The authors thank J. Anderson, N. Leprovost, M. Thompson,
 D. Tsiklauri and A. Zinober for useful comments. We also thank J. Douglas for numerous helpful discussions.

\begin{table}
\begin{center}
\begin{tabular}{lll}
\hline
\hline
$\overline\gamma \rightarrow \infty$&No resonance&$\langle n'^2\rangle\propto\overline\omega^{-2},\propto\overline\gamma^{-1}$ \\
 & & $\langle n'v_x\rangle\propto\overline\omega^{-2},\propto\overline\gamma^{-1}$\\

\hline
$1<\overline\gamma\ll\infty$&  $\overline\omega\gg1$ &$\langle n'v_x\rangle\propto\overline\omega^{-1.96}$\\
&                       &$\langle n'^2\rangle\propto\overline\omega^{-1.901}$\\&          $\Omega> 1$&$\langle  n'v_x\rangle\propto\Omega^{-1.015}$\\
&                       &$\langle  n'^2\rangle\propto\Omega^{-1.402}$\\
&           Doppler &$\langle  n'v_x\rangle\propto\overline\omega^{-0.486}\propto\Omega^{-0.534}$\\
&          resonance   &$\langle n'^2\rangle\propto\overline\omega^{-0.476}\propto\alpha^{-0.533}$\\
\hline
$0<\overline\gamma\ll 1$ &Parametric  & \\
& \& Doppler resonance&\\
\hline\hline
\end{tabular}
\caption{Summary of results with $0<\overline\gamma\ll 1$, $1<\overline\gamma\ll\infty$ and $\overline\gamma\rightarrow\infty$ describing the long, medium and short correlation times respectively.}
\end{center}
\end{table}
%








\end{document}